\documentclass[twocolumn,showpacs,preprintnumbers,amsmath,amssymb]{revtex4}
\usepackage{tabularx,graphicx}

\usepackage{color}
\usepackage{hyperref}
\hypersetup{
    colorlinks=true,
    linkcolor=blue,
    filecolor=blue,      
    urlcolor=blue,
}

\usepackage{color}

\usepackage{ulem}   

\usepackage{soul}

\begin{document}

\newcommand{\beq}{\begin{equation}}
\newcommand{\eeq}{\end{equation}}
\newcommand{\beqn}{\begin{eqnarray}}
\newcommand{\eeqn}{\end{eqnarray}}
\newcommand{\bmath}{\begin{subequations}}
\newcommand{\emath}{\end{subequations}}
\newcommand{\bra}[1]{\langle #1|}
\newcommand{\ket}[1]{|#1\rangle}

\def\correspondingauthor{\footnote{Corresponding author: jhirsch@ucsd.edu}}

\title{Clear evidence against  superconductivity in    hydrides  under high pressure}

\author{J. E. Hirsch$^{a*}$  and F. Marsiglio$^{b}$ }
\address{$^{a}$Department of Physics, University of California, San Diego,
La Jolla, CA 92093-0319\\
$^*$Corresponding author. Email: jhirsch@ucsd.edu\\
$^{b}$Department of Physics, University of Alberta, Edmonton,
Alberta, Canada T6G 2E1}

\begin{abstract} 
The Meissner effect, magnetic field expulsion, is a hallmark of superconductivity. 
Associated with it, superconductors exclude applied magnetic fields.
Recently Minkov et al. presented experimental results  reportedly showing
{\it ``definitive evidence of the Meissner effect''} in sulfur hydride and lanthanum hydride under high pressure \cite{e2021,e2021p},
and   Eremets et al. argued that ``{\it the arguments against superconductivity (in hydrides) can be either refuted or
explained}'' \cite{e2022}.
Instead,  we show here that   the evidence presented in those papers does not support the case for superconductivity in these materials.
Together with experimental evidence discussed in earlier papers, we argue that this strongly suggests
  that hydrides under pressure  are not high temperature superconductors.
\end{abstract}
\pacs{}
\maketitle 

\section{introduction}
The era of high temperature superconductivity in hydrides under high pressure was spawned by  the reported
discovery of superconductivity in sulfur hydride in 2015 by Eremets and coworkers \cite{sh3,sh32014},  with critical temperature 203 K, higher than any other critical
temperature known before. Since then, it has been reported that superconductivity at high temperatures
occurs also in 11 other hydrides under pressure \cite{materials,materials2,materials3}. These experimental works are strongly motivated and guided by
theoretical predictions of superconductivity in these materials based on the conventional
BCS-electron-phonon theory of superconductivity \cite{marscarb,review,review2,roadmap}.

Instead, we have recently argued that  the experimental evidence presented so far does not provide conclusive proof
of superconductivity in any of these 
materials      \cite{hm,hm2,hm3,hm4,hm5,hmnature,hm6,eumine,huangmine,oncshchi,preprint1,epl}. 
Others have also questioned experimental \cite{dc,talantsev,mazov}  and theoretical \cite{qtheory1,qtheory2} evidence for
superconductivity in some of  these materials. Therefore, the very  existence of high temperature
superconductivity in hydrides under pressure is now in doubt \cite{mao}.

In the initial paper~\cite{sh3}, Eremets and coworkers presented magnetic evidence of superconductivity
in sulfur hydride, that was questioned in ref.~\cite{hm4}. Recently, Minkov and coworkers provided 
new magnetic evidence for superconductivity in sulfur hydride as well as in lanthanum hydride \cite{e2021,e2021p}, and argued
that it provides definitive evidence for superconductivity. Eremets et al. provided further arguments and
experimental results in ref. \cite{e2022}.
On the contrary, we argue in this paper that these new measurements together with the old measurements provide
convincing evidence {\it against} the existence of superconductivity in these hydrides, in our opinion.

\section{comparison of old and new magnetic evidence for superconductivity in  sulfur hydride}
To have confidence that experimental results reflect true physics of the material  being studied, it is essential
that measurements are reproducible, not only within one lab and experimental group but also in different settings.
Unfortunately, no other group has reported measurements of magnetization in sulfur hydride 
(nor any other hydride) under high pressure.

The experimental results on sulfur hydride reported by the Eremets group in 2015 \cite{sh3} and 2021 \cite{e2021,e2021p}
report critical temperatures of $203$~K and $196$~K respectively, i.e. they are very close to one another. The size of the samples
used is also reported to be similar, approximately $80 \ \mu m$ \cite{sh3} and $85 \ \mu m$ \cite{e2021p} in diameter and
``a few $\mu m$'' in thickness  in \cite{sh3} and between $2.1 \ \mu m$ and $3.1 \ \mu m$ in thickness in \cite{e2021p}.
However, the measured magnetic moment of the samples under an applied magnetic field of the same
magnitude  {\it differ by a factor of 5}, with the 
2015 sample having the larger magnetic moment.  No explanation for this large discrepancy is given in \cite{e2021p}.
These magnetic moments are measured not by field cooling but by applying a magnetic field to an already cold
sample, and for sufficiently small fields so  that the field should not penetrate the sample, hence such large differences are not expected for samples of similar sizes.

More importantly, the magnitude of the lower critical field reported in 2015 \cite{sh3} was 30~mT, 
the same quantity in 2021 was reported to be 1.9~T \cite{e2021} and subsequently 0.82~T \cite{e2021p}, that is a difference  of a factor of 60 and
27 respectively from the value 7 years ago. 

The magnitude of the London penetration depth was reported to be 125~nm in 2015 \cite{sh3} and
between 18 and 23 nm   in 2021 \cite{e2021p}, 
a difference of a factor of  5. 

Thus, for the past 7 years the physics community was asked to believe that sulfur hydride had been
proven to be a 203~K superconductor based on magnetic measurements that were wrong by these very significant
factors. The magnetic evidence reported by Eremets and coworkers in 2015 \cite{sh3} was regarded by many
to be the strongest evidence that hydrides under pressure are high temperature superconductors. 
Now, Eremets et al. themselves are telling us that that evidence was flawed \cite{e2021p}. In other words, the entire edifice of high temperature superconductivity in pressurized hydrides was
built on dubious  foundations. 

\begin{figure} [t]
 \resizebox{8.5cm}{!}{\includegraphics[width=6cm]{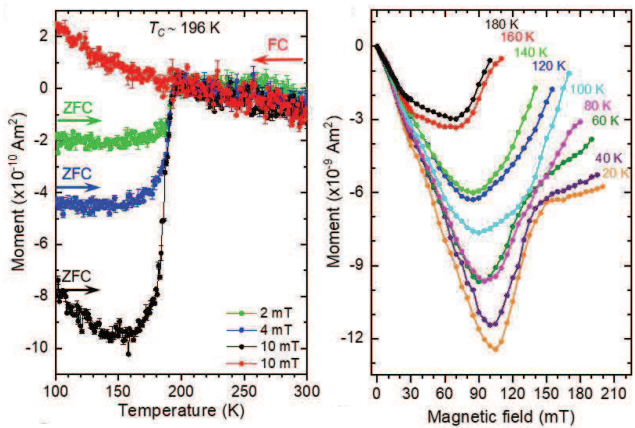}} 
\caption{ Left panel: magnetization versus temperature of sulfur hydride reported in ref.~\cite{e2021p},
under field cooling (FC) and zero field cooling (ZFC).
Right panel: magnetization versus applied magnetic field for sulfur hydride reported in ref.~\cite{e2021p}.  Reproduced under a CC BY 4.0 License  from ref. \cite{e2021p}.}
\label{fig3}
\end{figure}

\section{magnetic evidence in 2021}

It is also remarkable that after the reported magnetization measurements in 2015 using a
specially designed miniature nonmagnetic DAC cell  
that could accommodate a SQUID magnetometer \cite{sh3},
no new experimental results using that  sophisticated apparatus and technique  were reported for a full 6 years, neither by the
authors of ref.~\cite{sh3} nor by anybody else. Yet during those years, 
about 30 new reports of high temperature superconductivity in 12 different pressurized hydrides were 
published \cite{materials,materials2,materials3}.

Fast forward to 2021 \cite{e2021,e2021p}. Figure~1 shows on the left panel the reported magnetization versus temperature under an applied magnetic field,
and on the right panel the magnetization versus magnetic field for various values of the temperature \cite{e2021p}. 
It can be seen that no evidence of a superconducting transition is seen under field cooling 
on the left panel. While this behavior has been observed in some strongly type II superconductors \cite{nusran18,ni},
it has never been observed for type I or weakly type II superconductors, to our knowledge.
According to ref.~\cite{e2021p}, this material is a weakly type II superconductor,
with Ginzburg-Landau parameter $\kappa=12$. 
The reported London penetration depth is remarkably small, $22 nm$, indicating that
the material has a large superfluid density and small degree of disorder.
Such materials always exhibit a robust Meissner effect, i.e. magnetic flux {\it expulsion}.
We argue that the fact that this material does not show {\it any} evidence of magnetic field
expulsion under  field cooling, the signature of
superconductivity, is clear and direct evidence that the material is not a superconductor.

\begin{figure} [t]
 \resizebox{8.5cm}{!}{\includegraphics[width=6cm]{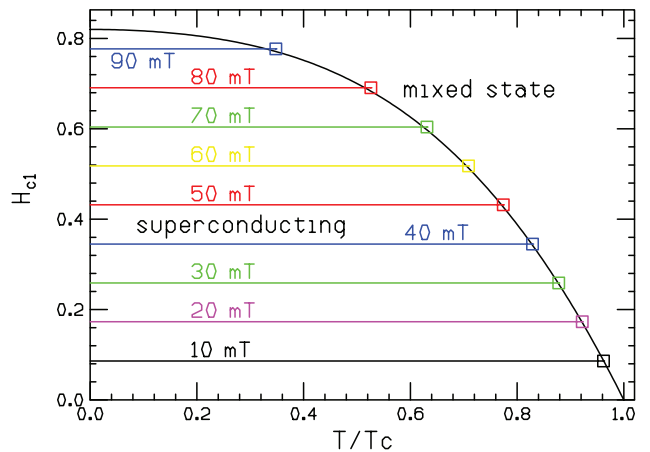}} 
\caption{ Lower critical field versus temperature for a standard superconductor according to Ginzburg-Landau theory.
The zero temperature critical field is $0.82$~T. The numerical values of the field given in mT indicate the
applied field $H_p$, that becomes $H_{c1}=8.5 H_p$ due to the demagnetization factor
$1/(1-N)$ for $N=0.88$ reported in ref.~\cite{e2021p}. }
\label{fig3}
\end{figure}

Ref.~\cite{e2021p} claimed that the sample is a flat disk of demagnetization factor $N=0.88$. From this information, and  from the
observation that a magnetic field at low temperatures of magnitude $H_p\sim 96\ mT$ starts to penetrate
the sample, it follows  that 
the lower critical field of sulfur hydride is $H_{c1}(T=0)=0.82 T$. In figure~2 we plot the behavior of the lower critical field of a standard superconductor as a
function of temperature inferred from Ginzburg-Landau theory. The horizontal lines indicate the values
of the lower critical field $H_{c1}$ for the values of applied field indicated, $H_p=10 \ mT, \ 20 \ mT$, etc.
The squares indicate the critical points for each value of the applied magnetic field.

\begin{figure} [t]
 \resizebox{8.5cm}{!}{\includegraphics[width=6cm]{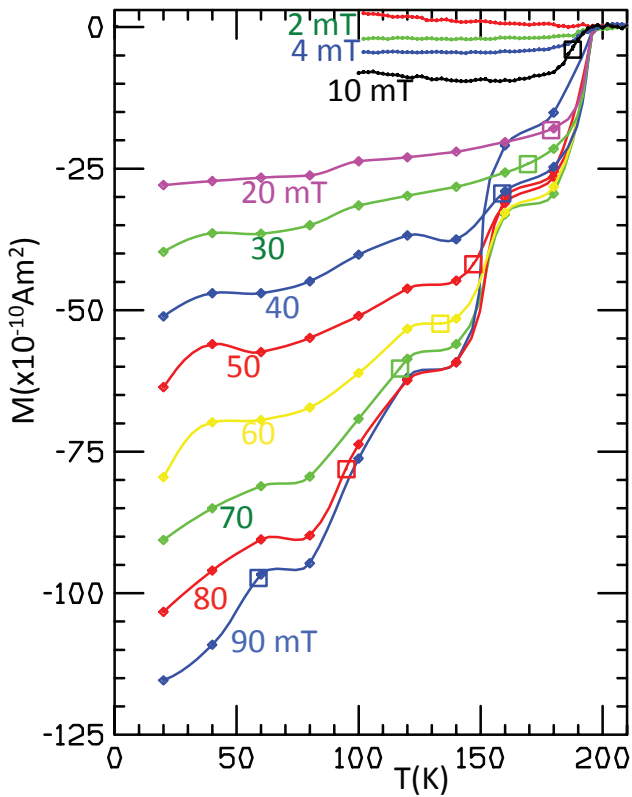}} 
\caption{ Magnetic moment versus temperature inferred from the experimental results of ref.~\cite{e2021p}
shown in Fig.~1, for various values of the applied magnetic field $H_p$ given in $mT$ in the figure. The colored squares indicate the point on the curve
for the applied magnetic field that corresponds to the critical temperature given by the phase boundary in Fig.~2.
}
\label{fig3}
\end{figure}

For each value of the magnetic field, for temperatures lower than that critical value, the magnetic field should be completely excluded from the sample,
except for the small region within $\lambda_L$ of the surface. If the penetration depth is   $22\ nm$ and
the sample diameter and height are $85 \ \mu m$ and $2.8 \ \mu m$ respectively, this implies
that $99\%$    of the sample remains field free except very close to the critical temperature. 
 Therefore,
we expect the magnetization versus temperature to be essentially flat below each critical temperature.
This is approximately consistent with what is seen for the three curves shown in the left panel of Fig.~1.
For those values of the applied field, $H_p=2 \ mT, 4 \ mT$ and $10 \ mT$, the critical temperatures are 
$194.4$~K, $192.8$~K and $187.8$~K, all close to the zero field critical temperature $196$~K.

From the reported values of magnetic moment versus magnetic field shown in the right panel of fig.~1 we can
extract the behavior of magnetic moment versus temperature for applied fields larger  than the values shown
on the left panel of fig.~1. This is shown in fig.~3. The value of the abscissa for each square is the critical temperature for
each value of the applied field inferred from fig.~2, and  we would expect the magnetic moment
to be approximately $constant$ for temperatures lower than those values, since the magnetic field is completely
excluded from the sample. The behavior seen in fig.~3 is {\it qualitatively different} from this expectation.

Indeed, it can be seen in fig.~3 that the sample is unaware of the fact that
it underwent a superconducting transition at its critical temperature, denoted by the square on each curve. The 
magnetization in each case (except for the selected ones that the authors of ref.~\cite{e2021p} 
chose to show us on the left panel of
fig.~1, also reproduced in fig.~3), 
continues its downward trend as the temperature is lowered further,
oblivious to the fact that the magnetic field is no longer in its interior.
This could be interpreted as revealing yet another new property of nonstandard superconductors \cite{hm2} not shared by
standard conventional or unconventional superconductors: an ability
to change the magnetic field in their surroundings while keeping the magnetic field in their interior
constant (i.e. equal to zero).

In the foregoing we have assumed the value of $H_{c1}=0.82 T$ for sulfur hydride inferred by Minkov et al.  in ref.  \cite{e2021p} from 
the magnetization data shown on the right panel of Fig. 1.  They used the widely accepted criterion that $H_{c1}$ is the value
of the applied field where the magnetization versus $H$ starts to deviate from linearity \cite{bean62} due to vortex penetration
into the superconductor, and an estimated demagnization factor $N\sim 0.88$ ($1/(1-N)=8.5$). 
There are however a variety of reasons that could change  the value $H_{c1}=0.82 T$, as discussed in the following.

(a) The demagnetization factor is difficult to estimate in the absence of precise information about the sample's dimensions,
and could be larger or smaller than the value $N=0.88$ assumed by Minkov  et al  \cite{e2021p}. In ref. \cite{e2021},
the demagnetization factor was assumed to be larger,  $N=0.95$, and a larger critical field value $H_{c1}=1.9 T$ was estimated.
In Ref. \cite{e2021p}  an estimated range of possible values $0.74 T<H_{c1}<1.09 T$ was given.
 In ref. \cite{e2022} it was estimated that 
for the sulfur hydride sample studied in 2015 the
 demagnetization factor  was $N=0.88$  and 
 a value $H_{c1}=0.55T$ for that sample was inferred from the measured magnetization values, substantially larger than the value $H_{c1}=0.03T$ originally inferred in 2015 yet
quite a bit smaller than the $H_{c1}=0.82 T$ assumed by Minkov et al in    \cite{e2021p}.


(b) It is generally believed that the Bean-Livingston barrier \cite{bl}, arising from the attractive image force on vortices near
the surface, may
prevent vortex penetration for a range of fields  above  $H_{c1}$, in which case the magnetization continues to be linear for fields
larger than $H_{c1}$. For example, Joseph and Tomasch \cite{joseph} 
find that in as-deposited and annealed films of Pb-Tl alloys the linear behavior of magnetization versus H can 
persist up to almost three times the bulk $H_{c1}$. On the other hand, for a FeSe single crystal 
Abdel-Hafied et al. find \cite{abdel} that the Bean-Livingston barrier does not play a role. The Bean-Livingston barrier is believed
to be particularly important for samples with very smooth surface boundaries and large values of the
Ginzburg-Landau parameter $\kappa$. Neither of those criteria appear to be likely to be important for the
cases under consideration here.

(c) In the presence of strong flux pinning, the deviation from linearity in the magnetization versus field curve may be
very gradual and hence the location of $H_{c1}$ could be underestimated, as discussed by Naito et al. \cite{naito} and
Reedyk et al. \cite{reedyk}.
This may contribute to the fairly wide spread in $H_{c1}$ values inferred for high $T_c$ cuprates.
For example, in Fig. 5a of ref. \cite{reedyk} the deviation from linearity does not become apparent up to above
a factor 1.5 higher than the estimated $H_{c1}$.

\section{Enormous $H_{c1}$ and density of states}

In ref. \cite{hm3}, in connection with an analysis of a nuclear resonant scattering experiment on sulfur 
hydride \cite{nrs}, we pointed out that a lower critical field $H_{c1}\sim 2.6\ T$, a thermodynamic critical
field $H_c\sim 10.6\ T$, and a London penetration depth $\lambda_L=10.0 \ nm$ are completely incompatible with the physics of standard superconductors. 
Those values are remarkably close to the values  $H_{c1}=1.9$~T, $H_c=9.8$~T and $\lambda_L=12.7$~nm inferred in Ref.  \cite{e2021}.

Taking into account the range of values of parameters suggested in Refs. \cite{e2021p}, \cite{e2021} and \cite{e2022},  let us assume for example that due to uncertainty in the
dimensions of the sample the demagnetizing factor is $N=0.8$ instead of the $N=0.88$ \cite{e2021p} or $N=0.95$  assumed by Minkov et al
in Refs. \cite{e2021p,e2021}.
This then yields $H_{c1}=0.48$~T, $H_c=4.1$~T and $\lambda_L=20.4$~nm as an alternative to the values assumed
by Minkov et al \cite{e2021,e2021p}, with an $H_{c1}$ even lower than  the range $0.74-1.09T$ estimated in \cite{e2021p}, thus allowing for the possibility that
some of the other effects discussed at the end of the previous section could artificially increase the apparent value of $H_{c1}$.

Let us start by computing the critical current. 
For the value $H_{c1}=1.9$~T estimated for sulfur hydride in ref.~\cite{e2021},  a magnetic field of magnitude smaller than
$1.9$~T should be completely excluded from the interior of a long cylinder. 
The current density circulating near the
surface is, according to London's equation
\beq
J_c=\frac{c}{4\pi \lambda_L}H_{c1}
\eeq
which for the above given values of $H_{c1}$ and $\lambda_L$  yields 
\beq
J_c=1.19\times 10^{10}Amp /cm^2. 
\eeq
This value is at least  two to   three orders of magnitude larger than critical currents
of any other known superconductor, whether type I or type II, hard or soft \cite{talanjc}. 
If instead we use our alternative values for the parameters assuming $N=0.8$ given above, the critical current is
$J_c=1.23\times 10^{9}Amp /cm^2$, still much larger than for any other known superconductor.

Instead, 
refs. \cite{e2021p} and \cite{e2021} estimated a value of the critical current density
\beq
J_c\sim 7\times 10^6 Amp /cm^2
\eeq 
using magnetization measurements and the Bean model. However, such an estimate can only be valid for a strongly type II superconductor with much smaller $H_{c1}$ and larger $\lambda_L$, such that the value of magnetic field given by the
expression $H=(4\pi \lambda_L/c)J_c$ is larger than $H_{c1}$. In such a case 
 the magnetic field penetrates the sample and 
generates vortices that are pinned near the sample surface, and the strength of the pinning potential determines
the critical current according to the Bean model. That calculation is not applicable for the case at hand here to determine the critical current, since $H=(4\pi \lambda_L/c)J_c$, with $J_c$ given by Eq. (3), is much smaller than $H_{c1}$.

The density of states at the Fermi energy $g(\epsilon_F)$ can be obtained from the standard theromodynamic relation
\beq
\frac{H_c^2(0)}{8\pi}=\frac{1}{2}g(\epsilon_F)\Delta^2
\eeq
with $\Delta$ the energy gap at zero temperature. From the standard BCS relation 
$2\Delta/k_BT_c=3.53$ and $T_c=196$~K \cite{e2021} we have $\Delta=29.8$~meV,
and using $H_c=9.8$~T yields
\beq
g(\epsilon_F)=\frac{0.537 \ states}{spin-eV A^3}.
\eeq
This is an enormous density of states. 
For comparison, using density functional  theory the density of states
of sulfur hydride was estimated to be  0.019 states/(spin-eV $\AA^3$) \cite{review2}, twenty eight times smaller.  
Our alternative values for the parameters assuming $N=0.8$  yield a density of states that is smaller than Eq. (5) by a factor
of six, so more than four times larger than the expected value.

In ref.~\cite{hm3} we discussed in more detail why such numbers are  incompatible with standard
superconductivity. Barring a qualitatively different superconducting state unlike that of all known 
superconductors, this implies, in our opinion, that the properties measured in ref.~\cite{e2021} interpreted to
``unambiguously confirm superconductivity'' in fact indicate that the material that was measured
was not a superconductor.

The analysis discussed above for sulfur hydride applies equally well to the measurements for
$LaH_{10}$ reported in ref.~\cite{e2021p}. It is claimed that for $LaH_{10}$ the zero temperature 
lower critical field, thermodynamic critical field and London penetration depth are
$H_{c1}=0.55$~T, $H_c=5.1$~T and $\lambda_L=30$~nm \cite{e2021p}, comparable to the values reported for
$H_3S$ ($H_{c1}=0.82$~T, $H_c=5.5$~T, $\lambda_L=22$~nm)and equally impossible. 


\section{sample quality}
The samples were prepared by pulsed laser heating of a precursor containing $NH_3BH_3$ as a source of hydrogen and either
$S$ or $LaH_3$. Then the precursor was heated by traversing a $5 \mu m$ laser spot horizontally and vertically across the sample.

The fact that the laser spot is much smaller than the estimated diameters of the samples suggests that the resulting sample cannot possibly
be a single crystal. Instead, there are likely to be many different regions of the sample of size of order of the laser spot that are connected
by weak links (V. Struzhkin, private communication (2021)) that would also give rise to pinning centers that could explain the absence of magnetic flux expulsion upon 
cooling. It would be beneficial for a better understanding if a detailed mapping of the variations of structure within the sample could be performed (see, e.g. 
Ref.~\onlinecite{diffr}).

\section{a smoking gun?}
Adding to the arguments presented in the previous sections, we argue that Fig. S1 of the paper \cite{e2021p} is a smoking gun
that provides clear evidence for  the faulty analysis and conclusions of ref. \cite{e2021p}.

\begin{figure} [t]
    \resizebox{8.5cm}{!}{\includegraphics[width=6cm]{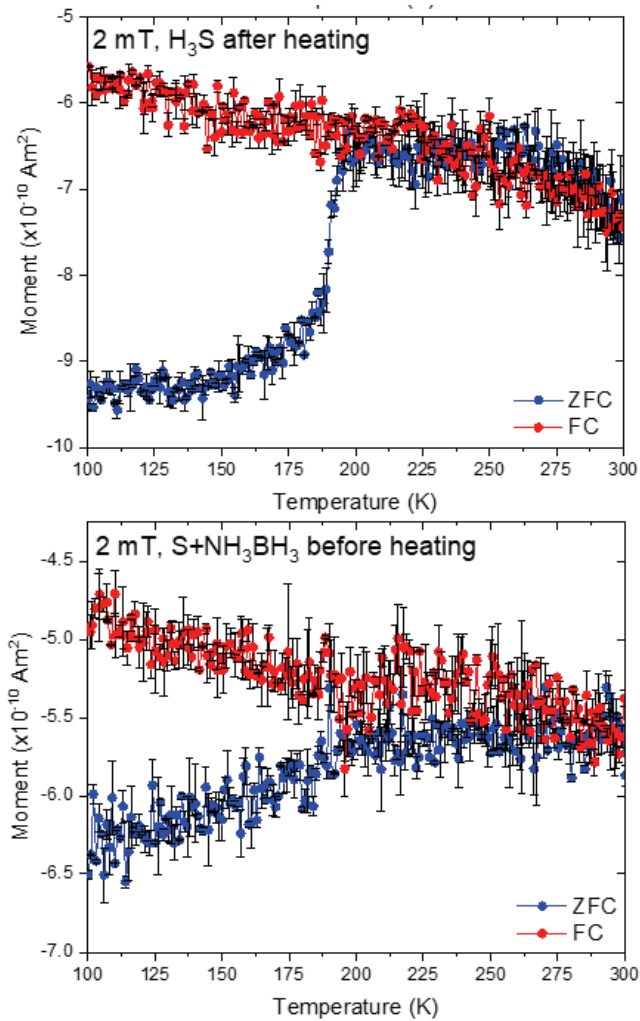}} 
\caption{Middle left (top panel here) and lower left (bottom panel here)
of Fig. S1 of ref. \cite{e2021p}. 
The bottom panel corresponds to the precursor sample, and the top
panel to the assumed superconducting sample. See text for discussion. 
Reproduced under a CC BY 4.0 License  from ref. \cite{e2021p}}
\label{fig3}
\end{figure}

\begin{figure*} [t]
 \resizebox{15.5cm}{!}{\includegraphics[width=6cm]{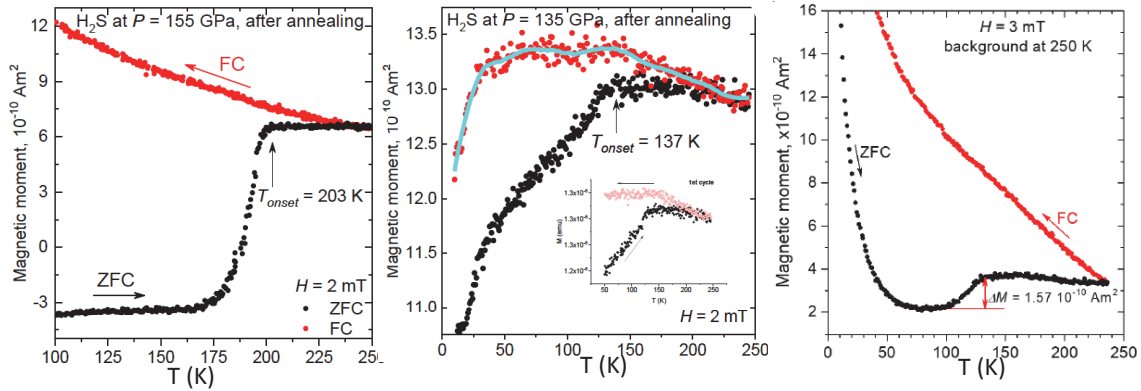}} 
\caption{  Magnetic moment versus temperature for sulfur hydride under field cooled (red points) and zero field cooled (black points) conditions
for three different runs performed in 2015, reported   in ref. \cite{sh3} in 2015 and in ref. \cite{e2022}, fig. 5 (a) (left panel), fig. 5 (b) (center panel) 
and fig. 6 (a) (right panel), reproduced under the Creative Commons CC BY license  from ref. \cite{e2022}.}
\label{fig3}
\end{figure*}

In Fig. 4, the bottom panel shows the measured magnetic moment versus temperature
under an applied magnetic field of 2 mT for the
sample precursors, namely $S+NH_3BH_3$ before undergoing the laser heating process that generates the
supposedly superconducting sample. The zero field cooling and field cooling curves approximately coincide for
temperatures above 200K and diverge below 200K, with the ZFC magnetic moment decreasing and
the FC moment increasing as the temperature is further lowered. Precisely the same behavior,
attributed to superconductivity, is
observed in the sample after heating, shown in the top panel
of Fig. 4 (and in the left panel of Fig.~1, red and green curves, after subtraction of the background signal from the diamond anvil cell).

{\it How does the precursor sample know that the critical temperature of the superconducting sample will be close to 200K?}
Why does an applied magnetic field give rise to a different magnetization for the non-superconducting sample under
field cooling and zero field cooling, also seen in the lower middle and right panels of Fig. S1 for other values of
the applied magnetic field?

The authors attribute the difference in the FC and ZFC curves shown in fig. S1 to ``contamination by magnetic
pieces'' (ref. \cite{e2021p} and V. Minkov, private communication). 
Clearly such effects, which will not necessarily have the same temperature dependence
before and after heating, and hence cannot be simply subtracted out, could also be responsible for the 
signals interpreted  as superconductivity.

We argue that the observed behavior in the lower panels of Fig. S1 of ref. \cite{e2021p},
which obviously is not indicative
of  superconductivity,
but clearly is due to experimental artifacts or properties of the background, strongly suggests that the similar divergence of FC and ZFC curves seen for the samples   after heating, interpreted as due to
superconductivity, could be also  caused by  the same  experimental artifacts or properties of the background.

\section{Eremets et al.  2022:  \textbf{\textit{`experimental evidence and details'}} }

In reference \cite{e2022}, Eremets and coworkers describe in greater detail the experiments that in their view reveal high
temperature superconductivity in hydrides under high pressure. Here we address several of the points made in that paper
posted in January 2022, more than 7 years after the announced discovery of  high temperature
superconductivity in sulfur hydride \cite{sh3,sh32014}.

In the introduction of \cite{e2022}, the authors argue  that nuclear resonant scattering experiments \cite{nrs}, infrared spectroscopy \cite{timusk} and 
ac susceptibility measurements \cite{huang} have supported superconductivity.
We have provided elsewhere detailed arguments for why those experiments were flawed \cite{hm3,hm5,hm6,huangmine}.
The authors of \cite{e2022} ignore  the issues that we raised in those references. In particular, they state 
in connection with ref. \cite{nrs} that
{\it ``it appears that the nuclear resonance scattering is a new, non-trivial, and sophisticated technique to detect superconductivity. Hopefully it will find further use to study novel near room-temperature superconductors in difficult conditions, such as ultra-high pressure, and, perhaps in compounds exhibiting superconductivity even above room temperature.''}
We point out that it  is peculiar that in the ensuing 6 years since ref. \cite{nrs} was published  not a single experiment using this technique was reported,
neither for a hydride nor for any other suspected or confirmed superconductor. Nor was that technique used to detect the Meissner
effect of any superconductor ever before the experiment reported in ref. \cite{nrs}.

Fig. 5 shows measurements of the magnetic moment of sulfur hydride versus temperature by the
Eremets group around the year 2015. The left panel was published in the 2015 paper \cite{sh3},
the center and right panel were unpublished until 2022 \cite{e2022}.
The left panel suggests a transition to superconductivity, and the magnetic moment changes from positive to negative
as expected for a superconducting transition. Instead, in the center and right panels the magnetic moment
does not change sign. 
Additionally, the magnitude of the drop in magnetic moment is a factor of 10 smaller on the right panel compared to the left panel.
For the center panel, the drop in magnetic moment occurs over a temperature interval of about 100 K, versus a temperature
interval of about 20-30 K for the left and right panels. Over that interval, the drop in magnetic moment in the center
panel is a factor of 5 smaller than in the left panel.

Why are these three panels so different from each other? The magnitude of the magnetic moment drop 
should only depend on sample dimensions, not temperature or pressure, and the sample dimensions for the
three cases were presumably similar. The width of the transition should be similar in different experiments but it is not.

The fact that there are all these differences between the three panels suggests that the behavior observed is not
due to superconductivity. 
Amongst the three, the left panel, published in 2015,  is the one that shows something closer to what is expected for a superconducting transition.



The strongest evidence for magnetic field {\it expulsion} under field cooling, i.e. Meissner effect proper, is provided by
the authors of \cite{e2022} in their Fig. 5h. In Fig. 6 we reproduce their Fig. 5h in the upper panel, and in the lower panel we
show the same figure with the blue line drawn by hand removed. We argue that {\it ``The subtle step(s) observed on FC curves''}
that according to the authors of \cite{e2022} is   indicated by the blue line  in the upper panel is no longer apparent in the lower panel when
the blue line and the arrow are removed.

\begin{figure} [t]
 \resizebox{8.5cm}{!}{\includegraphics[width=6cm]{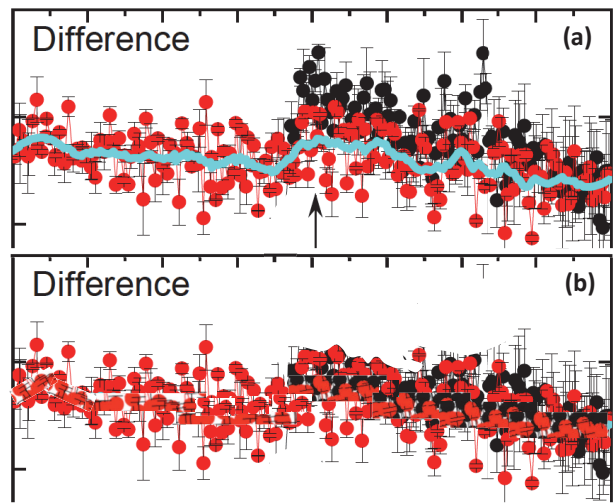}} 
\caption{Upper panel: alleged signature of the Meissner effect upon field cooling indicated by the blue line, from ref. \cite{e2022} Fig. 5h,
reproduced under the Creative Commons CC BY license  from ref. \cite{e2022}.
Lower panel: same picture with the blue line removed, i.e. we simply replaced the blue line by less distracting  red points and
white background that now
hide the blue line. We  also removed some distracting black points. All remaining black points should be ignored --- they
correspond to zero field cooled measurements. There is no apparent transition as a function of temperature indicated by the red points
in the lower panel. }
\label{fig3}
\end{figure}

Regarding sample dimensions, 
Eremets  et al.  in 2022 \cite{e2022}  start in the introduction by informing  readers that {\it ``the typical size of samples  residing in a diamond anvil cell (DAC) is necessarlly
small - of the order of $\sim 50 \mu m$ in lateral dimensions''}.  The word {\it ``necessarily''} in this context is  misleading.  

The maximum pressure that can be achieved in a diamond anvil cell is a function of the culet dimensions \cite{maoreview}. This is
discussed for example in ref. \cite {culets} and references therein. In their 2015 paper \cite{sh3},
Eremets et al. reportedly used {\it ``a culet of $40-80 \mu m$'' }and reported measurements in the range
100 - 200 GPa. In their 2021 paper \cite{e2021p}, Eremets et al. reported using culet sizes of $\sim 75$ and
$\sim 90 \mu m$ and pressures up to 167 GPa were applied.

While for the higher pressures the smaller size culet may be necessary,
superconductivity was reportedly detected in ref. \cite{sh3} at around 120 GPa. 
Such a pressure can be easily achieved with a $200 \mu m$ culet and possibly even larger \cite{culets}.
Such culets should allow for sample diameters much larger than the $\sim 50 \mu m$ that Eremets et al. tells us
in 2022 are the sizes of the {\it``tiny samples''} that necessarily have to be used. 
If the samples were  not so ``tiny'' but instead closer to 200 $\mu m$ in diameter or even larger and correspondingly
thicker, it should be much easier to unambiguously demonstrate  a magnetic signature of superconductivity if it existed. 

Regarding other measurements, particularly resistance, the 2022 paper by Eremets et al.   \cite{e2022} states:
{\it ``$H_3S$ was independently synthesized and superconductivity was fully confirmed [12-16]''}. We remark that:

 (i) Ref.  [12] \cite{12}  (2018)  is coauthored by Eremets, so it hardly qualifies as `independent'. The single resistance curve shown there
showing a drop starting at  temperature above 150 K has  a  width of more than 70 K, significantly larger than the resistive transitions shown in ref. \cite{sh3}
of width 10 K or less.

(ii) Ref.  [13] \cite{13} (2020)  is not coauthored by Eremets but the single experimental graph indicating superconductivity is the same
resistance measurement reported in ref. [12]  \cite{12}.

(iii) Ref.  [15] \cite{15} (2019)  is not coauthored by Eremets, it reports a single experimental curve of resistance versus temperature showing
a $15 \%$ drop in resistance at $200$ K, and the same curve with {\it ``subtraction of residual resistance''}, i.e. subtracting $85 \%$ of the
resistance, showing that it now drops (not surprisingly)  to zero.

(iv) Ref.  [14] \cite{14} (2019) shows the same experimental curve (with residual resistance subtracted) as ref. [15] \cite{15}.

(v) Ref.  [16] \cite{huang} are the ac susceptibility measurements that ref. \cite{huangmine} showed to be flawed.

We argue that the above  calls the statement of {\it ``fully confirmed''} made in ref. \cite{e2022} based on its 
references [12-16]  into question.

Finally, we would like to comment on the implication of the following paragraph of Eremets et al. 2022 paper \cite{e2022} reproduced below in full:

{\it ``For a long time we did not clearly appreciate the role played  (by) the quality and integrity of the samples. Even when the sample has a large enough size and clearly shows superconductivity in electrical transport measurements, the magnetic susceptibility signal can turn out to be elusive or below the sensitivity of the SQUID magnetometer. The reason for that can be the granular or non-uniform distribution of the superconducting phase in samples. The electrical current finds a continuous path through superconducting grains and metallic grain boundaries in the transport measurements whereas much smaller thin superconducting grains have a relatively small superconducting volume leading to a smaller signal due to the demagnetization factor.''}

The authors explicitly acknowledge that current circulating through {\it ``metallic grain boundaries''} 
providing {\it ``a continuous path''} can be interpreted as   showing superconductivity in their experiments.
If their superconducting grains have a {\it ``relatively small superconducting volume''} compared to the total volume it
implies that a significant part of the path through which the current circulates is metallic and not superconducting.
If the current necessarily has to go through metallic grain boundaries, which necessarily have non-zero resistance,
the material cannot sustain a persistent current and cannot possess macroscopic phase coherence. 
We argue that a material that cannot exhibit these defining features of superconductivity cannot be called
a superconductor, irrespective of how small its resistance is or how much its resistance depends on the ionic mass.

\section{a litmus test for hot hydride superconductivity}
We have proposed in ref. \cite{hm5} that detection of trapped flux would provide convincing evidence for the existence of
persistent currents in these materials. It would appear that to do this test would not be more difficult than to perform
the measurements reported in refs. \cite{e2021p} and \cite{sh3}.

To rationalize the complete absence of flux expulsion upon field cooling seen in the left panel of fig. 1, one could hypothesize that
there is a large concentration of  defects that trap the magnetic field and prevent  it from being expelled, even at the large cost
of condensation energy implied by the very large $H_c$. Let us entertain that possibility. 
If after  the field cooling process shown on the left panel of Fig. 1, for applied  magnetic field smaller than $H_{c1}$,  the applied   field is turned off,
those defects  should prevent the interior magnetic field from decaying and 
a remnant magnetization should be detectable for  several hours, days or months  thereafter. As shown in refs. \cite{ft1,ft2},
the remnant magnetic moment should be given approximately by the difference in the FC and ZFC moments for the
same magnetic field. 

So far there have not been experimental reports that trapped flux, persisting for a long time, has been detected in hydrides. 
This is in contrast to what happened with the cuprates, where from the very beginning it was seen that samples that did
not strongly expel magnetic fields trapped magnetic fields and this was considered independent evidence for
superconductivity \cite{muller}. The current situation for hydrides
is depicted in Fig. 7. Standard superconductors, conventional and unconventional, are described by the
extreme behaviors shown in the first and second row of Fig. 7, as well as behavior intermediate between them, i.e. part of the flux is expelled upon cooling and the part
that is not expelled remains trapped after the external field is removed.
Thus far hydride superconductors, and non-superconductors, are described by
the third row: none of the flux is expelled upon field cooling, and none is trapped when the external field is removed.

\begin{figure} [t]
 \resizebox{8.5cm}{!}{\includegraphics[width=6cm]{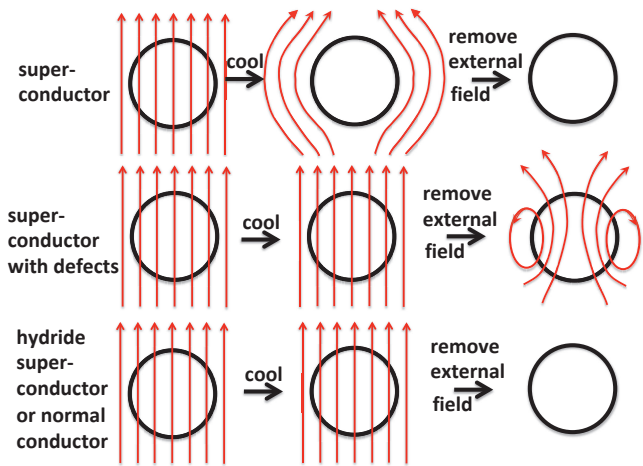}} 
\caption{ First two rows: expected behavior of standard superconductors. The first row shows complete flux
expulsion, as occurs in clean type I superconductors or type II superconductors with weak pinning centers. The 
second row shows small or zero Meissner fraction, as occurs in superconductors with  many defects/strong pinning centers that
trap magnetic field. When the external field is removed, the trapped field remains. Of course intermediate behavior
between the first and second row is possible and in fact most common, namely some of the flux is expelled and some
is trapped when the external field is removed. The third row shows
the behavior of normal metals and of hydride superconductors (as far as we know to date): they neither expel magnetic field nor
trap magnetic field.}
\label{fig3}
\end{figure}

If  such a trapped flux is detected for a hydride sample in the future, $and$ it is not detected when the same process is performed for the precursor sample
before heating, it will provide convincing evidence that persistent currents flow in the material in the absence of applied magnetic field.
Then the unlikely possibility that these materials are `nonstandard superconductors' \cite{hm2,hm3} will have to be seriously considered.
Conversely, if no field trapping is observed, it will provide strong confirmation of our arguments that indicate that these systems
are not superconducting.

Similarly we have suggested \cite{hm5} that in a nuclear resonant scattering experiment (NRS) \cite{nrs} that claimed
to show flux exclusion from sulfur hydride from the absence of quantum beats in the NRS spectra under an applied 
magnetic field, it would be straightforward to verify the presence of trapped field by showing that quantum beats
are present after field cooling and removal of the external field.

\section{summary and conclusions}

The magnetic measurements reported in refs.~\cite{e2021,e2021p,e2022} were intended to establish that hydrides under
pressure are indeed high temperature superconductors.  Ref. \cite{e2021} claimed to present  {\it unambiguous} (three times)
and {\it definitive} (once) evidence for  high temperature superconductivity in sulfur hydride and lanthanum hydride under pressure.
Ref. \cite{e2022} claims to have presented {\it ``very solid evidence for high-temperature superconductivity in hydrogen-rich compounds under high pressures''}. 
In this paper we argued that instead these recent experimental papers \cite{e2021p,e2021,e2022}  not only have  $not$ established that these hydrides are superconductors but rather they have increased the likelihood that they are not.
The papers provide
a snapshot of where the field is today, more than 7 years after the reported discovery of this new class of superconducting materials. Compare this snapshot with a snapshot of where the field of $any$ other new class of superconducting materials
that withstood the test of time was 7 years after its discovery, particularly those with higher $T_c$. It is precisely because only a few
researchers can perform these difficult   high-pressure
experiments  that the onus is on    them  to provide convincing evidence for superconductivity.

%

 Let us summarize our conclusions.

(1) The measurements show that if the materials are superconductors,
they have critical current values that are one to three orders of magnitude larger than those of all standard superconductors.

(2) Their lower critical field and thermodynamic critical field are between one and two orders of magnitude larger
than for any standard superconductor.

(3) Their density of states is between one and two orders of magnitude larger than what is expected for
a material with such composition.

(4) Their London penetration depth is  smaller than the smallest penetration depths found for any other material  
including very pure elements.

Conclusions (1) to (4) are based on our assumption, consistent with the authors' conclusions \cite{e2021,e2021p,e2022} inferred
from the magnetization data, with which we agree, that the lower critical field in these
materials, if they are superconductors, is of
order 0.5 T or larger. If instead the critical field was one to two orders of magnitude smaller, these conclusions would be invalidated.
We do not believe that this is plausible, given the magnetic evidence discussed.

(5) These materials show absolutely no evidence for magnetic field expulsion in the presence of 
a magnetic field, as seen in figs.~2c~and~2d and the 9 panels in fig. S1 of ref.~\cite{e2021p}, as well as figs. 5 and 6a in ref. \cite{e2022}.
This is completely unprecedented for weakly type II superconductors with such small values of the London
penetration depth that imply very high values for the superfluid densities.

(6) The sample preparation process indicates that the resulting samples are not single crystals and have  weak links and disorder,
making even more implausible  the claimed large values of lower critical field and critical current, and low value of the London penetration depth.

(7) The measured magnetic moment versus temperature under zero field cooling conditions, except for very
small magnetic field values, shows
monotonic behavior of approximately constant slope (Fig. 3)  with absolutely no signature of 
a superconducting transition at the expected values of the critical temperature.

(8) The measured magnetic moment interpreted as due to superconductivity is five times smaller than
the magnetic moment interpreted as due to superconductivity in the 2015 paper  \cite{sh3} for a 
sample of similar size.

(9) The measured magnetic moment versus temperature in the presence of a small magnetic field for 
precursor nonsuperconducting samples shows similar behavior to the behavior allegedly signaling superconductivity
in the alleged superconducting samples.

(10) Detection of trapped flux  \cite{hm5} that persists for long periods after field cooling and removal of the external field
has not been reported so far. If it is found, it would show that the materials can carry persistent currents, hence it
is a superconductor. This is assuming that contributions to magnetic remanence from magnetic sources can be
ruled out, e.g. by comparison with the behavior of untreated samples not expected to be superconducting. If 
trapped flux is looked for and not found, its absence would confirm that
the materials are not superconductors.

It should also be pointed out that Ref. \cite{e2021p} states in the caption of Fig. 2 {\it ``Expulsion of magnetic field by the superconducting $Im-3m-H_2S$
and $Fm-3m-LaH_{10}$ phases ...''} when in fact the figure shows no evidence for magnetic field expulsion.
Ref.~\cite{e2021} misleadingly claimed that ``the Meissner effect'' was demonstrated,
even in the title. The fact is, the Meissner effect is magnetic field {\it expulsion}, not magnetic field {\it exclusion}. 
Magnetic field exclusion was known to researchers in 1911, magnetic field expulsion was only discovered in 1933.
Refs.~\cite{e2021p}, \cite{e2021} and \cite{e2022}  showed zero evidence for magnetic field expulsion. Similarly ref. \cite{nrs} claimed to report
``direct observation of the expulsion of the magnetic field'' from sulfur hydride, when in fact no field cooling was even performed.


The experiments reported by Minkov et al. \cite{e2021p} and Eremets et al. \cite{e2022} are performed under very difficult experimental conditions, 
at very large pressures and with very small samples, and a variety of effects unrelated to the physics of the sample
but instead due to properties of the background or the experimental apparatus could come into play. Under such challenging conditions
we suggest that it is very important to guard against confirmation bias, the tendency to pay undue attention to
observed features that confirm prior beliefs and ignore others  that don't. 
 While we recognize that it may be very difficult to obtain
clear evidence of a Meissner effect in a very small sample, we point out that as discussed in ref. \cite{hm5},
a rather large remnant magnetic moment should remain trapped in the samples after field-cooling in a large
magnetic field if one is to believe the
nuclear resonant scattering experiment \cite{nrs,pro}; no such
observation has been reported so far.
We don't know what could be the origin of the features found in the experiments that are attributed to be
superconductivity by Eremets et al., but have argued in this paper that in our opinion they are inconsistent
with superconductivity.

In other recent papers we have analyzed various other reported evidence for superconductivity in pressurized hydrides
and concluded that every experiment was flawed, namely:

(1) ac magnetic susceptibility reported to show a superconducting transition in sulfur hydride \cite{huang} was
shown to be due to an experimental artifact \cite{huangmine}.

(2) Optical reflectance measurements that reportedly showed that sulfur hydride  is  a superconductor \cite{timusk}
were shown to be flawed \cite{hm6}.

(3) The reported observation of a Meissner effect in sulfur hydride using nuclear resonant scattering \cite{nrs}
was shown to be flawed \cite{hm3}.

(4) Magnetic susceptibility measurements for a room temperature superconducting hydride \cite{roomt} were
shown to be flawed \cite{oncshchi,preprint1,epl,dirk}.

(5) Magnetic susceptibility measurements reported for lanthanum hydride showed weak 
and very broad peaks \cite{struzhkin},
inconsistent with the width of the presumed transition shown in resistivity measurements
of that material \cite{lah,lah2,hm4}.

(6) Resistance measurements for a room temperature superconducting hydride \cite{roomt} and several 
other hydrides \cite{eremetsy,sh3,lah2,semenoklay}
were shown to be anomalously sharp, and/or with an unchanging width in the presence of an 
applied magnetic field, indicating that the drops in resistance
were not due to superconductivity \cite{hm,hm2,hmnature}.


To extract $H_{c1}$ from magnetization data is a notoriously difficult task \cite{naito,gutierrez,mosh}, and as discussed at the end of Sect. III, other effects not considered in refs. \cite{e2021p,e2022} could play a role to reduce the value of $H_{c1}$ somewhat. 
However, we believe it is extremely unlikely that such a reduction would be of the magnitude necessary 
 to invalidate our conclusions in this paper, namely one to two orders of magnitude.

The field of high temperature superconductivity in hydrides was launched in 2015 by the publication of 
ref.~\cite{sh3} by Eremets and coworkers. Now, the same author and coworkers present evidence
\cite{e2021,e2021p,e2022}  that (a) invalidates
the evidence for superconductivity presented in the 2015 paper and (b) is in itself
 flawed, as discussed in this paper. From this, together with our analysis of the totality of magnetic evidence 
and other experimental evidence for superconductivity in hydrides discussed in our earlier papers, we argue that the 
most likely conclusion is that hydrides under pressure are not high temperature superconductors.
Whatever
the origin of their anomalous behavior is, whether intrinsic or due to experimental artifacts or both, in our opinion it is not
due to superconductivity.

\begin{acknowledgments}
We are grateful to D. Semenok for calling ref.~\cite{e2021} to our attention, to
V. S. Minkov, S. L. Bud'ko and M. I. Eremets for clarifying correspondence on their papers, and to V. Struzhkin
 and R. Prozorov for discussions.
JEH would like to thank S. Shylin for extensive discussions and sharing of information on refs. \cite{sh3} and \cite{e2022}.
FM  was supported in part by the Natural Sciences and Engineering
Research Council of Canada (NSERC), and by an
MIF from the Province of Alberta. 

\end{acknowledgments}

\end{document}